\documentclass{article}

\usepackage{arxiv}

\usepackage[utf8]{inputenc} % allow utf-8 input
\usepackage[T1]{fontenc}    % use 8-bit T1 fonts
\usepackage{hyperref}       % hyperlinks
\usepackage{url}            % simple URL typesetting
\usepackage{booktabs}       % professional-quality tables
\usepackage{amsfonts}       % blackboard math symbols
\usepackage{nicefrac}       % compact symbols for 1/2, etc.
\usepackage{microtype}      % microtypography
\usepackage{lipsum}

\usepackage{adjustbox}
\usepackage{amssymb}
\usepackage{lineno}
\usepackage{graphicx}
\usepackage{lineno}
\usepackage{graphicx}
\usepackage{cuted}% for environment `strip`
\usepackage{caption}
\usepackage{hhline}
\usepackage{float}
\usepackage{multirow}
\usepackage{amsmath}
\usepackage[labelsep=period]{caption}  % to use one dot rather than two for figures

\usepackage{mathdots}
\usepackage{mathtools}
\usepackage{nicefrac}
\usepackage{placeins}
\usepackage{booktabs}
\usepackage{lipsum}
\usepackage{blindtext}
\usepackage[euler]{textgreek}
\usepackage[linesnumbered,ruled,vlined,noend]{algorithm2e}
\usepackage[usenames, dvipsnames]{color}
\usepackage{booktabs,lipsum,makecell}

\title{Compressive coded rotating mirror camera for high-speed imaging}

\author{
  Amir Matin\\
  Institute of physics and quantum science\\
  Heriot Watt University\\
  Edinburgh, United Kingdom \\
  \texttt{am626@hw.ac.uk} \\
   \And
  Xu Wang \\
  Institute of physics and quantum science\\
  Heriot Watt University\\
  Edinburgh, United Kingdom \\
  \texttt{x.wang@hw.ac.uk} 
}

\begin{document}
\maketitle

\begin{abstract}

We develop novel compressive coded rotating mirror (CCRM) camera to capture events at high frame rates in passive mode with a compact instrument design at the fraction of the cost compared to other high-speed imaging cameras. Operation of CCRM camera is based on the amplitude optical encoding (grey scale) and a continuous frame sweep across a low-cost detector using a motorized rotating mirror system which can achieve single pixel shift between adjacent frames. Amplitude encoding and continuous frame overlapping enable the CCRM camera to achieve high number of captured frames and high temporal resolution without making sacrifices in the spatial resolution. Two sets of dynamic scenes have been captured at up to 120 Kfps frame rate in both monochrome and colored scales in the experimental demonstrations. The obtained heavily compressed data from the experiment are reconstructed using the optimization algorithm under the compressive sensing (CS) paradigm and the highest sequence depth of 1400 captured frames in single exposure has been achieved with the highest compression ratio of 368 compared to other CS-based high-speed imaging technologies. Under similar conditions CCRM camera is 700$\times$ faster than conventional rotating mirror based imaging devices and could reach frame rate of up to 20 Gfps. 

\end{abstract}

% keywords can be removed
\keywords{High-speed imaging \and optical encoding \and Compressive sensing (CS)}

\section{Introduction}
\par  High-speed imaging has shown an exceptional potential in capturing ultrafast transient phenomena in a variety of applications such as screening the physiological processes in biological tissues \cite{1}-\cite{4} high-throughput blood cells screening \cite{5},\cite{6}, fluorescence confocal and lifetime microscopy \cite{7}.\cite{8} which require cameras with capture rates between Kfps to Mfps. In the fields of biomedical research and clinical applications for example, high-speed imaging allows the detection and tracking of cells, plasma, and other molecules of interest in a specimen individually or as a group with high sensitivity and precision. Bulky design, high cost and complex operation of the conventional high-speed imaging systems make them almost inaccessible in technical environments especially in the resource-limited areas. As a result, the development of portable, low cost and low maintenance high-speed imaging systems is particularly important for such environments.  

\par At the early development stages of high-speed cameras, the rotating mirror camera \cite{9} was one of the first commercially available instruments to achieve 25 Mfps. The principle of this scheme relied on the rotation of a single mirror that directed the incident light (frames) towards a film strip and was replaced by the array of Charged-Coupled Devices (CCD) in the later development stages (e.g. Brandaris) \cite{10}. Alongside the aforementioned rotation mirror high-speed cameras, there are several other commercially available devices (such as IX and Phantom cameras) that can operate at several hundred Kfps frame rate which can be suitable for a range of applications such as combustion studies, that is the investigation of the chemical reaction between the substances, schlieren imaging (a non-invasive method to visualize density gradients within otherwise invisible flows), microfluidics, particle image velocimetry, material tensile testing and many more \cite{11}-\cite{14}. The operation of these instruments rely on the high-speed electronics such as the custom-made charge-coupled devices (CCD) and complementary metal oxide semiconductor (CMOS) detectors that can achieve such high frame rates that are extremely costly to build and demand for high storage capacities and high maintenance costs. The necessity of observing phenomena at high frame rates have driven the imaging society to push the boundaries of the imaging technologies to achieve superior technologies compared to their previous generations and achieve Gfps and up to Pfps frame rates where they have proven their significance in various applications such as observing the dynamics of laser-induced breakdowns \cite{15}, observing the 3D shapes of the hidden objects using ultrafast time-of-flight imaging techniques \cite{16}, non-scanning tomographic multispectral imaging methods that enables single-shot cross-sectional image acquisition at femtosecond time scale \cite{17}, ultrafast spectroscopic videography of dynamic non-repetitive events \cite{18} and three-dimensional high-speed imaging \cite{19},\cite{20}.

\par Current imaging techniques however are associated with their unique disadvantages such as the dependency on the precise repetition of the ultrafast event during the captures (multi-shot imaging), lacking the capability of imaging the luminescent transient events, monochrome scaled captures, low number of captured frames (short duration of recording), demanding storage and transmission capacity requirements, extremely high built costs, high maintenance, oversized dimensions and highly complex operations. As a result, overcoming these shortcomings while achieving high acquisition rates in the system becomes an essential requirement for majority of the affected applications. In the recent years, several novel ultrafast imaging technologies using highly dispersed ultrashort pulses have been developed that are capable of achieving  Kfps - Tfps  frame rates \cite{21}-\cite{24}. These active imaging methods use ultrashort optical pulses generated by the femtosecond laser source and various optical techniques to perform space-spectrum and spectrum-time mapping to capture the ultrafast phenomenon at up to Tfps acquisition rates. Since the spatial and temporal dispersion are all realised in optical domain in these technologies, the electrical and mechanical bottlenecks in the conventional high-speed imaging systems can be eliminated. However, these imaging technologies are also associated with series of drawbacks such as the requirement for expensive short-pulse laser (operation only in active mode), the imposed limitations on the amount of pulse dispersion to avoid overlapping of the adjacent pulses that is also known as data-mixing, faster frame rates at the expense of imaging contrast, spatial resolution and detection sensitivity as well as their requirements for high data transmission, processing and storage capacities to enable capturing of the ultrafast phenomenon at high capture rates and longer durations. Techniques using differential detection and dictionary learning by discrete cosine transform (DCT) have been proposed for data compression in the time-stretched imaging systems \cite{25},\cite{26}. Virtual time gating technique using parallel signal channels is also proposed to overcome the aforementioned constraints on the time-stretched imaging systems however, the amount of dispersion is still limited to avoid the heavy overlap between the signals as well the requirement for multiple groups of detectors for signal detection which can be a costly solution \cite{27}. Furthermore, improvements on the spatial resolution and the detection sensitivity of the conventional imaging systems were achieved by using techniques such as asymmetric-detection time-stretch optical microscopy delivering ultrafast high-contrast and label-free imaging solution \cite{28}.

\par Alongside the mentioned studies, Compressive Sensing (CS) method has also been introduced to the high-speed imaging systems where by using an optical encoder in the space-to-spectrum mapping stage of the imaging system, up to 100x faster frame rates were achieved \cite{29},\cite{30}. These encoders are also integrated in the systems in the forms of a translating coded aperture to compress the data in spatial and temporal domains \cite{31}, via an external pulse pattern generator to apply random binary sequence on the pulse \cite{32} as well as the use of digital mirror device (DMD) in their optical systems \cite{33},\cite{34} that are capable of achieving frame rates from 1 Kfps up to 100 Gfps. However, these imaging systems are also restricted by their distinctive limitations such as the confined capture durations, sacrificed spatial resolution that is imposed by the limitations of the streak camera for 2D imaging setups, requirements for the pulsed illuminations as well as the controlled amounts of shift between the adjacent overlapping frames that constraints the sequence depth of these imaging modalities. The range of achieved advancements from these CS enabled imaging systems is a strong indication of the versatility of the CS framework in high-speed imaging applications and a robust proof that CS based systems can improve the overall performance of the system by a substantial amount however in these previous studies, the high build and maintenance costs, bulky dimensions but more importantly the operation complexity and the restricted number of captured frames at a limited spatial resolution has hindered the scientific and industrial fields from employing these CS enabled high-speed imaging systems in their research and applications.

\par To address these matters, we propose a novel coded compressive rotating mirror (CCRM) camera that enables high-speed imaging of a transient phenomenon at a substantially lower build cost and operation complexity using the off-the-shelf components in a compact instrument design. Our approach implements amplitude encoding (grey scale) in a dramatically simplified rotating mirror camera setup where the continuous exposure of the dynamic scene on the rotating mirror and the constant frame sweep across the detector surface enables the CCRM camera to reach the lowest possible distance between the sweeping overlapped frames that is a single pixel on the detector. This principle avoids the controlled positioning of the frames on the detector and allows the encoded frames to be uniformly overlapped within a single pixel of the adjacent ones and therefore reach the highest compression ratio and the highest number of recorded frames (sequence depth) in this type of imaging schemes \cite{29}-\cite{34}.  

\par CCRM camera differs from other CS based 2D imaging systems such as the proposed scheme at \cite{31}  where the binary encoder and the fully overlapped frames are replaced with a static encoder and pixel-shifted overlapping frames that enables the camera to capture much higher number of frames in both monochrome and colored formats and achieve higher reconstruction quality. CCRM also differs from the proposed scheme at \cite{34} where the requirements for the pulsed illumination from the target (scene) and the controlled amounts of shift between the adjacent frames are eliminated which allows the scene to be continuously captured from a natural self-luminant target or using a continuous external light source without any frame drops during the full exposure period of the detector. Furthermore, the recovery of the data using an amplitude encoder instead of a the binary format that is used in the previous studies, eliminates the binning process on the data hence significantly increasing the spatial resolution and the total number of overlapped and recovered frames compared to the aforementioned CS based imaging modalities. Additionally, the implementation of the calibration tracer blocks (discussed in the Principles section of this paper) provides a precise spatial location of the individual frames on the detector that enables the data reconstruction with higher fidelity. The heavily compressed data from the CCRM camera that are encoded by the optical encoder can only be recovered using that pattern which is considered as the “key” to the reconstruction algorithm. This property of the proposed scheme enables a highly secure and efficient data storage that utilizes the use of the CCRM imaging technology in the fields such as medical imaging and military based applications that require highly secured data handling and processing capabilities.

\section{Principles }

\par
\begin{figure}
\centering
\hspace*{-0.2in}
		\includegraphics[width = 25pc]{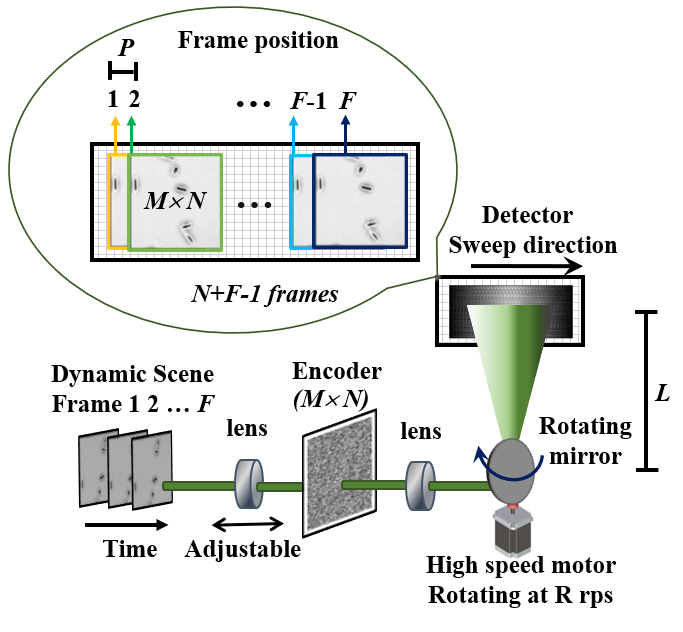}
		\captionsetup{justification=centering}
		\captionof{figure}{Configuration and operation principle of the proposed CCRM camera setup}   
		\label{fig:ccrm}
\end{figure}

 Depicted in Fig. \ref{fig:ccrm} is the configuration and operation principle of the proposed coded compressive rotating mirror (CCRM) imaging system. In the capturing stage, the image of the dynamic scene at time t is focused on a spatial encoder (with M rows and N columns), and then focused on the rotation mirror driven by a high speed motor that further reflects the image onto a 2 dimensional (2D) photo detector (such as CMOS or CCD). The detector module should have at least M rows and (N+F-1) columns in size with moderate readout speed. The rotation of this mirror continuously sweeps the individual frames across the surface of the detector module and overlaps them during a single exposure that creates a single pixel shift in-between the adjacent frames. Therefore, M$\times$(N+F-1)  pixels on the detector will record F of the coded frames at different times that are positioned within a single pixel distance from each other throughout the continuous illumination from the dynamic scene. Capturing a scene in a single exposure of the detector, eliminates the limitation of digitization and readout time of the camera from the proposed scheme. The mathematical representation of data acquisition process of the proposed CCRM imaging system can be formulated as

\begin{equation}
y = TCAx + n,
\label{eq:forward_model}
\end{equation}

where $y \in \mathbb{R}^{MN+(F-1)M \times 1} $ is the captured data by the detector in a vectorized format, $T  \in \mathbb{R}^{MN+(F-1)M  \times MNF} $ is the linear operator of frame shifting and overlapping that is built upon F identity matrices, $C \in \mathbb{R}^{MNF \times MNF}$ is the obtained motion profile of the sweep from the calibration points on the encoder in the form of a diagonal matrix, $A \in \mathbb{R}^{MNF \times MNF} $ is the matrix that holds F encoding pattern of $M \times N$ in a diagonal form, $x \in \mathbb{R}^{MNF \times 1} $ represents the original frames in a vectorized format, and $ n $  is the additive zero mean Gaussian noise.

\par Estimating x from y in Eq. \ref{eq:forward_model} is known as an ill-posed linear inverse problem (LIP) \cite{35} i.e., there is more than one feasible solution to this problem. The formulated sensing matrix referred to as TCA in Eq. \ref{eq:forward_model} enables an extremely high compression to be achieved on the observed temporally shifted and overlapped data however it is to be noted that this type of compression does not satisfy the Restricted Isometry Property (RIP) \cite{35} used in the general compressive sensing framework therefore, the data reconstruction is associated with the inevitable artefacts that is known as a lossy recovery. Many reconstruction methods such as dictionary learning based algorithms, Bayesian, Gaussian mixture models and maximum likelihoods have demonstrated their capabilities in solving such equations. Of these, we implement the Alternating Direction Method of Multipliers (ADMM) \cite{36} method that applies variable splitting to the cost function and solves the shaped Lagrange equations accordingly. This approach transforms the obtained equation into a minimization problem and solves the equation by minimizing the energy function via the repetitive calculations. Here, we select the total variation (TV) \cite{37},\cite{38} as the regularizer function where it has demonstrated high performance on various CS based algorithms \cite{39},\cite{40}. One of the advantages of using TV over the other regularizers is the edge preservation property that prevents hard smoothing of the edge features. This key characteristic averts the spatial information from merging with the background features and therefore preventing the loss of the critical information such as the boundaries and intensity amplitudes per pixels that are essential is applications such as high throughput cell screening where the cell counting the exact shape of the individual cells are the defining factors in the analysis.

Image reconstruction of the original scene can be achieved by solving the minimization problem that is

\begin{equation} 
\hat{x}= \arg\min_x{\frac{1}{2}\left|\left|y-TCAx\right|\right|_2^2+\rho_kD\left(x,\rho_{tv},\ w_{tv}\right)}
\label{eq:var_rec}
\end{equation} 

where $\rho_k\ ,\ \rho_{tv}$ are the variable regularization and denoiser threshold parameters that are adjusted based on the calculated error at each iteration and $w_{tv}$ is the regularizer weight for each horizontal, vertical and temporal domains and $D$ is a regularization function that promotes sparsity in the dynamic scene. The linear operation of frame shifting and overlapping (represented by operator T) results in the compression of the data in CCRM camera. This data compression ratio can be expressed as:
 
\begin{align}
compression~ratio = \frac{N\ \times\ F}{N\ +\ F\ -\ 1}
\label{eq:comp}
\end{align}

and therefore, higher numbers of captured frames F (longer durations of captures) result in higher compression ratio.

\begin{figure}
\centering
\hspace*{-0.2in}
		\includegraphics[width = 42pc]{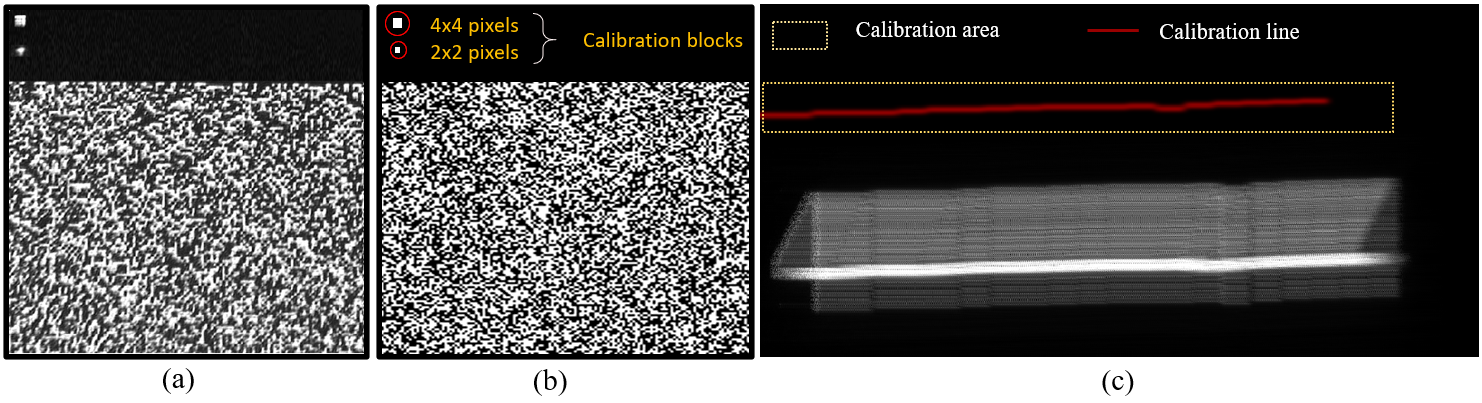}
		\captionsetup{justification=centering}
		\captionof{figure}{Comparison of the encoding patterns between the (a) observed pattern on the CMOS (b) image of the designed pattern. (c) Movement profile of the rotating mirror}   
		\label{fig:encoder}
\end{figure}

\par The obtained compression from the CCRM camera has the highest ratio compared to the other CS based imaging systems \cite{29}-\cite{34} that is attribute to the single pixel shift modality in the operation principle of the proposed imaging system that allows the overlap of higher number of consecutive frames in a single capture with high data reconstruction fidelity. Rotation of the motor is often associated with various types of flaws such as the backlash error, vibration at high speeds, design imperfections in the mirror holder attached to the motor etc. Hence, due to these inevitable mechanical errors, position of individual frames on the vertical axis of the detector may differ to the adjacent frames. On the other hand, Having a precise indication on the position of frames on the captured data is a necessity for the data reconstruction. Tracing these vertical movements are achieved by implementing the calibration blocks that are positioned on the encoder pattern in a designated location that leaves a distinct movement trace on the detector and is unique for each capture. Depicted in Fig. \ref{fig:encoder}(a-b), are the two designated calibration blocks at the top left corner of the encoding pattern with the dimensions of 2$\times$2 and 4$\times$4 pixels where they leave a trace of the movement on the detected image. These tracer lines are then extracted from the scanned data and used to define the exact position of each frame in the compressed scan. The 2$\times$2 block is used as a primary calibration block which is also used in the forward model and the reconstruction algorithm, however for the scenes with the lower amounts of illumination intensities on the CMOS, detecting the full calibration line with all the pixels present in the line can be a challenging task therefore, a larger 4$\times$4 block with higher light throughput assists the primary block in defining the position of the frames.

\par Fig. \ref{fig:encoder}(c) illustrates the scan of a static scene and marking of the boundaries in the selected area where the vertical shift in each frame is compared against the first (reference) frame. The primary calibration line color is marked as red in this picture for better visibility. The extracted motion profile is noted as C in Eq. \ref{eq:forward_model} that provides the vertical shift parameters to the reconstruction algorithm. Furthermore, as the rotation of the mirror generates the sequence of the overlapped frames on a cylindrically curved surface, the scene will encounter the so called “Petzval Field Curvature” aberration in which the paraxial sections in the image plane will appear slightly out of focus. The amount of blur on the frames will depend on the degree of rotation on the flat surface of the detector that is ~2 degrees in the CCRM camera and therefore the error produced by the filed curvature aberration is ~0.037mm which is negligible. This aberration however can be eliminated by selecting the central section of the field of view that effectively crops the blurred sections on the image plane.

\par In the optical setup of the CCRM camera, the dynamic scene is first collected by a 10$\times$ infinity-corrected microscopic objective (Thorlabs, RMS10$\times$) and directed towards the lens tube. The 2D image is then focused onto a static  encoder built in-house, with diverse pattern varieties and different spatial resolutions. The encoder patterns are printed on a soda-lime glass and have a 1:1 ratio between the blocked and transparent pixels. As the light passes through the optical encoder, the intensity values of the adjacent pixels interfere with the neighbouring pixels due to the light diffraction. This phenomenon can be moderately compensated by implementing the commonly used binning process (e.g. 2$\times$2 or 3$\times$3) on the detector \cite{20},\cite{33} and using larger encoding pixel dimensions on the encoder however this comes at a cost of the reduced spatial resolution of the detected image. To reduce the amount of interference between the adjacent pixels, encoding pixels with larger dimensions can be printed on the mask, however, this results in a lower encoding resolution on the scene, therefore having a negative effect on the recovery of the data. Our proposed new approach to tackle this problem in the CCRM camera is to take a reference image of the encoder pattern on the detector prior to recording of the dynamic scene. This detected pattern on the detector has a grey scale rather than binary values which is then used in the reconstruction algorithm.

\par The achieved frame rates ($F_r$) from the CCRM camera setup is calculated by the following expression

\begin{equation}
\label{eq:frate}
\begin{split}
F_r \approx \frac{2\pi\ .\ \ R\ .\ \ L}{P}~(fps),\\
subject~to~P<<L 
\end{split}
\end{equation}

where $R$ is the rotation speed of the mirror (rounds per second), $L$ is the orthogonal distance between the mirror and the detector surface and P is the width of each pixel in the detector (the distance between adjacent frames). In the conventional rotating mirror camera systems \cite{9},\cite{10}, the frame rate could be calculated using similar formulas to Eq. \ref{eq:frate} which is inversely proportional to the distance between adjacent frames that is the width of the detector with the dimensions of at least N$\times$P.  In this regard, the proposed CCRM camera operates at N times faster frame rate comparing to the conventional rotation mirror camera systems with the same mirror rotation speed R and arm length L. Assuming N=500 and the gap in between the adjacent detectors to accommodate the series of detector modules in the alignment of the conventional rotation mirror camera system, the CCRM operates in at least 700 times faster frame rate than the conventional rotation mirror camera systems with the same R and L parameters. In the conventional imaging systems, the frame rate of a camera is determined by the exposure time of the detector \cite{9}-\cite{14} and therefore limited by the speed of the detector, while in the CCRM camera, the entire dynamic scene is captured during single exposure of the detector that makes the frame rate of the CCRM camera independent of the detector’s exposure time and enables the use of low speed detectors to achieve high frame rate. In the experiments, CCRM camera uses an ordinary off-the-shelf CMOS module with dimensions of 1216$\times$1936 pixels (5.86 $\mu$m per pixel) and a maximum frame rate of 47 fps to capture dynamic scenes at up to 120 Kfps frame rate. The electrical motor used for the rotation of the mirror is a low-cost unit with the maximum rotation speed of 0.5 rotations-per-second (rps). This setup with the aforementioned components can record 1400 frames with spatial dimensions of 350$\times$350 and 200$\times$500 pixels with the data compression ratios of 280 and 368 respectively. These frames could be reconstructed with high fidelity using the formulated minimization function stated at Eq. \ref{eq:var_rec}.

\section{Results}

\par High-speed imaging is an imperative tool in medical and bio-chemistry fields with extensive application in high throughput cell screening \cite{21},\cite{28},\cite{30},\cite{41} microparticle flow cytometry \cite{12},\cite{13},\cite{24},\cite{42} and many others. In microfluidics and lab-on-chip studies, the diverse range of imaging applications with distinct properties of the samples require the imaging setup to be adapted to that particular experiment with minimum effort. The adaptability of the CCRM camera setup becomes evident when the frame rate of the system can be easily controlled by adjusting the rotation speed of the motor without altering the optical system or using the CCRM camera for both types of active (light source required) or passive (self-luminant) applications without any modifications in the system.

\par This is considered as one of the fundamental advantages of CCRM camera where the versatility and adaptability in imaging various types of scenes for variety of applications with different requirements in size, imaging speed, illumination properties etc. enables the proposed imaging technology to capture the dynamic scenes without any compromises in the operation principle of the system.In this regard, we configure the CCRM camera to observe flowing droplet samples in the microfluidic chip from injecting two types of immiscible liquids, transparent oil and chemical dye through a motorised dispenser into the chip channels. In this experiment, a continuous white light source is used to illuminate the microfluidic chip where droplets have the flow rate of ~0.9 m/s in the channels. The dynamic scene is collected from the object (using a 10$\times$, 0.25-NA, 10.6mm WD objective lens) and focused onto the encoder. The secondary lens system then collects the encoded image and directs it towards the rotating mirror. Three separate captures are recorded at the rates of 3, 50 and 120 Kfps and reconstructed through the proposed algorithm.

\begin{figure}
\centering
\hspace*{-0.2in}
		\includegraphics[width = 30pc]{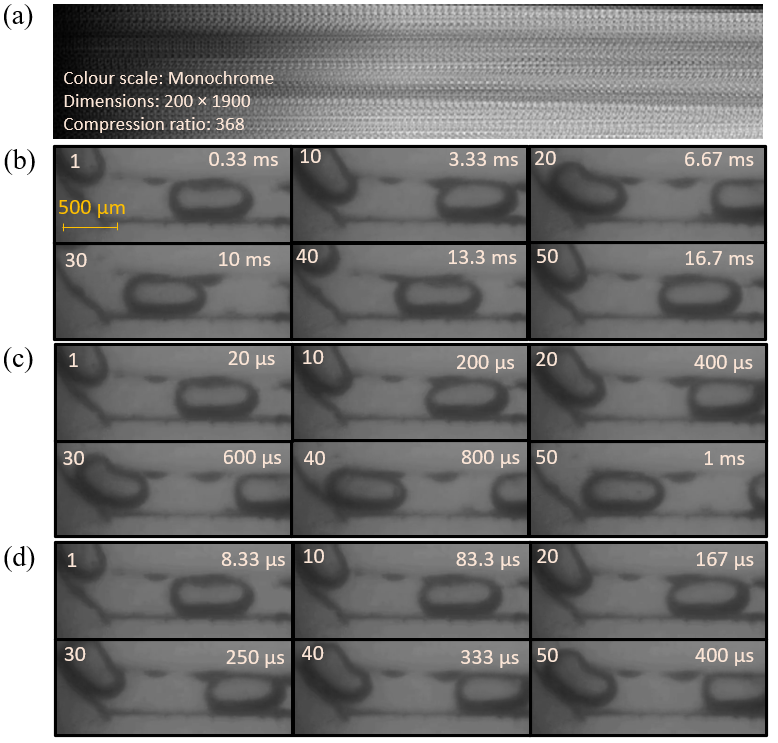}
		\captionsetup{justification=centering}
		\captionof{figure}{(a) Capture of the microfluidic chip at the frame rate of 120 Kfps and reconstructed frames from the captures at frame rates of  (b) 3, (b) 50 and (d) 120 Kfps. Frame numbers and time instance of the frames are shown in the top left and right corners of each frame respectively.}   
		\label{fig:fluidic_exp}
\end{figure}

\par Depicted in Fig. \ref{fig:encoder}(a) is a section of the encrypted and compressed captured data with dimensions of 200$\times$1900 pixels and the compression ratio of 368 (obtained from Eq. \ref{eq:comp})  for the droplets flow experiment at the frame rate of 120 Kfps.  Fig. \ref{fig:fluidic_exp}(b-d) illustrates some of the reconstructed frames obtained from solving Eq. \ref{eq:var_rec} by using the encoding key pattern (see Supplementary Movie 1 for the full reconstructed data captured at 120 Kfps). Achieved results demonstrate a high reconstruction quality with clear and distinguishable droplets flowing in the microfluidic chip with well-preserved edge features in the frames. Noticeable examples of this are the conserved bent edges for the first droplet at frames 30, 40 and 50 of the 120 Kfps capture dataset. Imaging of the droplets flow in the microfluidic chip at various frame rates demonstrates the capabilities of the CCRM imaging system in microfluidics and lab-on-chip studies where it can be easily adapted to record other bio-chemical phenomenon with their distinct properties and capture rate requirements.

\par In addition to the medical and bio-chemical applications, high-speed imaging has also been utilized in observing ultra-fast events in other fields such as laser ablation and plasma dynamics \cite{23}, fluorescence imaging \cite{33}, explosions and chemical reactions \cite{43}, surface inspection \cite{44} and many others. Particular example of this type of events that are continuously investigated in applied thermal engineering fields are so called “micro-explosions”. This phenomena can be observed in various scenarios such as electrically-triggered atomic emission spectroscopy (AES) \cite{45}, atomization process resulting in droplet disintegrations \cite{46} and laser-induced sample ignitions \cite{47}. Recording and observing these micro-explosions can help the investigators in gaining deeper understanding of the observed phenomenon. In general, these non-repetitive events that trigger a deformation process of a substance (gas, liquid etc) in micro-resolution that are associated with the so called “burst” process can be regarded as micro-explosions. An example of this type of event is the ignition of the gas lighter which uses an alloy of iron and cerium that emanates sparks during the rapid oxidization process where the dynamics of the scene share similarities with the aforementioned micro-explosion based applications. Here, in our next demonstration, we configure our experiment on the micro-explosion event in which the observed dynamic scene is the generated sparks from a handheld lighter module.

\par
\begin{figure}
\centering
\hspace*{-0.2in}
		\includegraphics[width = 28pc]{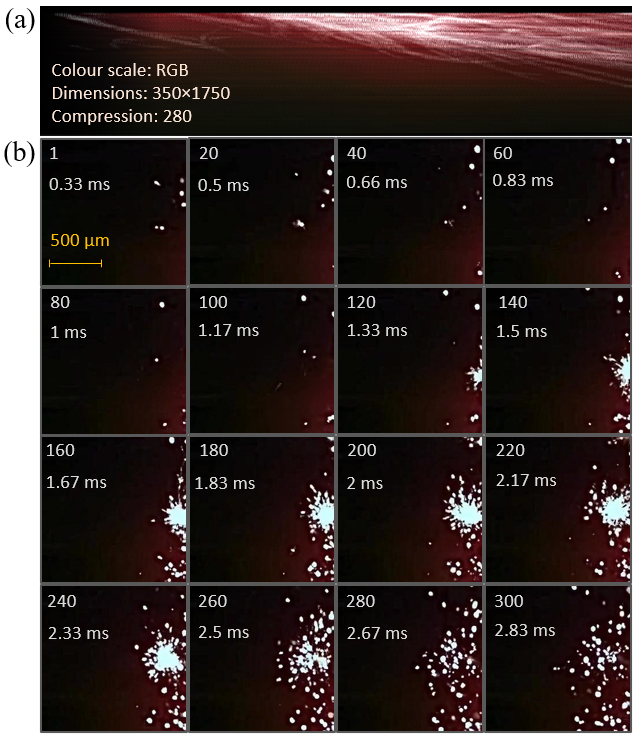}
		\captionsetup{justification=centering}
		\captionof{figure}{(a) Capture of the generated sparks from the lighter at 120 Kfps and (b) reconstructed frames from the Captures at frame rate of 120 Kfps. Frame numbers and the time instance of the frames are depicted in the top left corners of each frame.}   
		\label{fig:spark_exp}
\end{figure}

 Depicted in Fig. \ref{fig:spark_exp}(a,b) are the captured and the reconstructed data from movement of the generated sparks respectively (see Supplementary Movie 2 for the reconstructed data captured at 120 Kfps). As the dynamic of the scene has a 3 dimensional property, the random movements of the sparks occur in all spatial dimensions however the captured and reconstructed data are considered to be in a 2D format. As the light intensity varies in different sections of the device, higher amounts are more noticeable at the center of the spark groups.  

\par The proposed reconstruction algorithm has efficiently conserved the features of the event by preserving the sharp boundaries of the generated spark particles that are well distinguishable in the scene. Travel direction of the particles are random in the 3-dimensional spatial domain therefore majority of the particles are not positioned on the focus plane of the lens hence different intensity levels are observed per particle. To reconstruct this scene, Eq. \ref{eq:var_rec} has been extended into color domain where by separating the RGB channels, applying the algorithm on each individual channel and blending them back into single images, colored reconstruction of the frames can be achieved. The process of reconstructing the individual channels are decoupled form each other therefore they can be determined in a parallel fashion, hence the accelerated reconstruction can be achieved. 

\par Distinct properties in the conducted experiments demonstrate the flexibility of the  CCRM camera in capturing the high-speed dynamics of the scenes under diverse conditions with unique characteristics, such as

\begin{itemize}
  \item Illumination conditions, where in the first experiment (microfluidics), an external light source was used to provide the sufficient light for the scene to be observed whereas the second scene is a self-luminant transient phenomena that does not require the external source.
  \item Dynamics of the scenes, in which the first demonstration can be regarded as a controlled experiment where the scene is a repetitive event whereas the second scene has a random and non-repetitive nature.
  \item Imaging depth, where the first scene occurs in a 2 dimensional space where only the front surface (primary focus plane) of the chip is observed whereas for the latter experiment, the movement of the sparks happen in a 3D space.
  \item Colour properties, in which the first experiment was observed and reconstructed in monochrome whereas the second experiment was conducted in RGB format which can provide extra information about the compounds in the observed event.
\end{itemize}

\section{Conclusions }
CCRM is a novel passive high-speed imaging scheme based on amplitude encoding and continuous frame sweeping modality using inexpensive off-the-shelf electrical and optical components (motor, CMOS, and etc). This operation principle enables the CCRM camera to reach the physical limitation of such CS based imaging systems and achieve the minimum distance between the sweeping overlapped frames that is a single pixel on the detector without sacrificing the spatial resolution of the frames. CCRM camera eliminates the requirements for the pulsed illumination from the scene and controlled overlapping ratio between the adjacent frames and allows uniform and continuous sweep of the encoded frames across the detector and therefore reach the highest sequence depth of 1400 recorded frames with the compression ratio of 368 from an ordinary low-cost CMOS detector. The proposed CCRM camera is expected to have an immediate impact on the applications that are requiring high-speed, compact, low cost and easy to use imaging systems and has shown its capabilities in capturing dynamic events at high frame rates and long durations in monochrome and coloured scales. The replacement of multi-detector design of the conventional rotating mirror camera with a single detector in CCRM camera, considerably reduces the required rotation degree of the mirror and thanks to the continuous frame overlapping nature of CCRM camera, capturing rate of the system solely depends on the rotation speed of the mirror. CCRM camera can reach frame rates of up to ~20 Gfps with comparable rotating mirror mechanics used in the conventional rotating-mirror cameras. Achieved high compression ratios through the native built-in optical operation, reduces the requirement for storage and transmission capacities by significant amounts. CCRM camera holds a great potential for imaging fast dynamic phenomenon that can be adapted by a wide range of scientific and industrial fields. Alongside the demonstrated applications in bio-chemistry and physical science, CCRM camera can also be utilized as an effective optically-encrypted imaging system for high-speed applications in a diverse variety of applications such as medical and military based projects with highly sensitive captured data in which the full encryption is carried out during the scan prior to storing any information on storage units.

%%% Comment out this section when you \bibliography{references} is enabled.

\end{document}